\documentclass[epsfig,useAMS,usenatbib]{mn2e}
\usepackage{epsfig,graphics}
\newcommand{\be}{\begin{equation}}
\newcommand{\ee}{\end{equation}}
\title[Probing the Sagittarius stream with blue horizontal branch stars]
{Probing the Sagittarius stream with blue horizontal branch stars}
\author[Clewley \& Jarvis]{L. Clewley\thanks{clewley@astro.ox.ac.uk} \& Matt J.\,Jarvis\thanks{mjj@astro.ox.ac.uk}\\
Astrophysics, Department of Physics, Keble Road, Oxford, OX1 3RH, UK }
 
\date{Released 2002 Xxxxx XX}

\pagerange{\pageref{firstpage}--\pageref{lastpage}} \pubyear{2002}

\def\LaTeX{L\kern-.36em\raise.3ex\hbox{a}\kern-.15em
    T\kern-.1667em\lower.7ex\hbox{E}\kern-.125emX}

\begin{document}
\label{firstpage}
\maketitle
\begin{abstract}
We present 2-degree field spectroscopic observations of a sample of 96
A--type stars selected from the Sloan Digital Sky Survey Data Release
3 (SDSS DR3). Our aim is to identify blue horizontal branch (BHB)
stars in order to measure the kinematic properties of the tidal tails
of the Sagittarius dwarf spheroidal galaxy. We confine our attention
to the 44 classifiable stars with spectra of signal-to-noise ratio
$>15$\AA$^{-1}$.  Classification produces a sample of 29 BHB stars at
distances $5-47\,$kpc from the Sun. We split our sample into three
bins based on their distance. We find 10 of the 12 stars at
$14-25\,$kpc appear to have coherent, smoothly varying radial
velocities which are plausibly associated with old debris in the
Sagittarius tidal stream. Further observations along the orbit and at
greater distances are required to trace the full extent of this
structure on the sky. Three of our BHB stars in the direction of the
globular cluster Palomar (Pal) 5 appear to be in an overdensity but
are in the foreground of Pal 5. More observations are required around
this overdensity to establish any relation to Pal 5 and/or the Sgr
stream.  We emphasize observations of BHB stars have unlimited
potential for providing accurate velocity and distance information in
old distant halo streams and globular clusters alike. The next
generation multi-object spectrographs provide an excellent opportunity
to accurately trace the full extent of such structures.
\end{abstract}

\begin{keywords}
galaxies: individual (Sagittarius) -- Galaxy: halo  --   stars: horizontal branch -- Galaxy: structure -- Galaxy: stream
 \end{keywords}

\section{Introduction} 
There have been numerous searches for streams of material responsible
for the build-up of stellar halos in galaxies. In the Milky Way the
best studied is the Sagittarius (Sgr) dwarf spheroidal galaxy (Ibata,
Gilmore \& Irwin 1994) and its stellar stream (e.g. Majewski et
al. 2003). An extended stream of stars has also been uncovered in the
halo of the Andromeda galaxy (M31), revealing that it too is
cannibalizing a small companion (e.g. Lewis et al. 2004). Such streams
yield crucial information on the merging and accretion history of
galaxy halos and have been used to constrain the mass of the Milky Way
halo (e.g.  Johnston et al.  1999) and the mass of the halo in M31
(Ibata et al. 2004).

There are numerous studies reporting stellar streams associated with
the tidal debris of the Sgr system either trailing or leading it along
its orbit. The stars involved in the tidal disruptions can be used as
test particles which can lead to an understanding of the shape (Helmi
2004) and strength of the Milky Way potential. A number of studies
have sought to trace Sgr and its stream with a variety of stellar
populations. For example, Totten \& Irwin (1998) found evidence for
such streams using intrinsically rare, but very luminous, carbon
stars. The calibration of the carbon star distance scale remains
controversial and is further complicated by variability, dust and
metallicity effects. Numerous RR Lyrae stars have also been found to
be associated with the Sgr stream (Ivezi{\'c} et al. 2000; Vivas et
al. 2001; Vivas et al. 2005). The Quasar Equatorial Survey Team
(QUEST) survey (Vivas at al. 2004) found 85 RR Lyrae variables
covering about 36$^{\circ}$ in right ascension, of which 16 were
identified spectroscopically (Vivas et al. 2005). An all-sky view of
the Sgr stream has been investigated using M giants selected from the
Two Micron All-Sky Survey (2MASS) database (Majewski et al. 2003,
2004).

There is strength in diversity. Heterogeneous stellar populations not
only provide an important check on the distance and velocity
measurements of any putative stream but also probe different formation
epochs. Blue Horizontal Branch (BHB) stars provide yet another stellar
population ideal for exploring Halo structure (Sirko et al. 2004;
Brown et al., 2004; Clewley et al., 2005). Newberg at al. (2002) find
a distant sample of BHB stars that, they suggest, reside in the
trailing arm of the Sgr stream.  Further, Monaco et al. (2003) suggest
that a sample of BHB stars discovered in the core of the Sgr dwarf
spheroidal galaxy are the counterpart of the stars observed by Newberg
et al. (2002).

Both BHB stars and RR Lyraes are excellent standard candles, enabling
us to determine their distances to 5-10\%; by contrast the distances
to K and M giants are accurate to 20\% - 25\% (e.g. Dohm-Palmer et
al. 2001; Majewski et al. 2003). BHB and RR Lyrae stars are also old and
metal poor, so they are ideal for tracing old parts ($>$ 5 Gyr) of tidal
streams. The age of the Sgr M giants is controversial.  Majewski et
al. 2003 considered these stars to be younger than 5 Gyr with a
significant fraction of them only 2$-$3$\,$Gyrs old (Majewski et
al. 2003). This poses two problems for their use as tracers. First,
they may not be useful halo test particles as such dynamically young
tracers do not place very stringent constraints on the halo (Helmi
2004). Second, comparisons between the evolution timescales of the M
giants and the dynamical timescale required for the stars to populate
the stream suggest that the stars exist in a stream that took more
than their age to form. This inconsistency has been addressed by
Bellazzini et al. (2005) who finds the M giants were considerably older
($>$ 5 Gyr) than was originally thought and may even probe similar
epochs as RR Lyraes or BHB stars.

This paper is the first in a series that discusses the use of BHB
stars to directly trace the Sgr stream. BHB stars are A--type giants that are
easily identified in the Galactic halo as they lie blueward of the
main--sequence turnoff (e.g.  Yanny et al. 2000). Assembling clean
samples of remote BHB stars has been stymied by the existence of a
contaminating population of high--surface--gravity A--type stars, the
blue stragglers, that are around two magnitudes fainter.  Previous
analyses required high signal-to-noise ratio ($S/N$) spectroscopy to
reliably separate these populations (e.g.  Kinman, Suntzeff \& Kraft
1994), making identification of BHB stars in the distant halo
unfeasible. However, recently Clewley et al. (2002, 2004) developed
two classification methods that now enable us to overcome the
difficulties of cleanly separating BHB stars from blue stragglers.

In this paper we use these classification techniques, and the Sloan
Digital Sky Survey Data Release 3 (SDSS DR3) photometry, to isolate a
sample of BHB stars along the Sgr stream. In this new survey we target
stars along the equatorial strip in the region
$180^{\circ}<$RA$<249^{\circ}$, which are sensitive to the
different models that distinguish between halo flatness using the Sgr
stream (e.g. Mart{\'{\i}}nez-Delgado et al. 2004).

In Section 2 of this paper we describe the selection of the BHB
candidates from the SDSS data set, and outline the prescription for
transforming the SDSS $g,r$ magnitudes of A--type stars to $B,V$
magnitudes detailed in Clewley et al. (2005). Section 3 provides a
summary of the 2dF spectroscopic observations and data reduction
procedures. In Section 4 we classify these stars into categories BHB
and blue straggler. Section 5 presents the results of the
classification procedure and provides a summary table of distances and
radial velocities of the stars classified as BHB. Three BHB stars in
the direction of the globular cluster Pal 5 appear to be in an
overdensity but in the foreground of Pal 5; we compare our
measurements with those in the literature. In Section 6 we investigate
the evidence that the stars reside in a stream and are associated with
the Sgr tidal debris. Finally, in Section 7 we provide a summary of
the main conclusions of the paper. In this paper we use the coordinate
R$_{gal}$ to denote Galactocentric distances and the coordinate R$_{\odot}$ to
denote heliocentric distances.

\section{Selecting the BHB candidates}

\subsection{Colour selection}
Our aim was to use the SDSS photometric data to select a sample of
candidate faint BHB stars, with minimal contamination by quasi-stellar objects (QSOs) and
F--type stars. We selected the candidate BHB stars using the SDSS
point spread function (PSF) $ugr$ magnitudes of stellar objects. The
SDSS is 95\% complete for point sources to ($u, g, r, i, z$) = (22.0,
22.2, 22.2, 21.3, 20.5) and is saturated at about 14 magnitudes in $g,
r$ and $i$ and about 12 mag. in $u$. Consequently, we select stars
with 14.3 $< g <$ 19.5. The fainter limit was added in retrospect and
is a consequence of the 2dF instrument capability and the observing
conditions coupled with our strict signal-to-noise (S/N) requirements
for classification, which we outline below. All the SDSS magnitudes
discussed in this paper have been corrected for Galactic extinction,
using the map of Schlegel, Finkbeiner \& Davis (1998). We limited
ourselves to the equatorial stripe in the SDSS data set which is
observable from the southern hemisphere, with
$180^{\circ}<$ RA $<249^{\circ}$ (J2000).
\begin{figure}
\centering{
\scalebox{0.40}{
\includegraphics*[20,140][700,700]{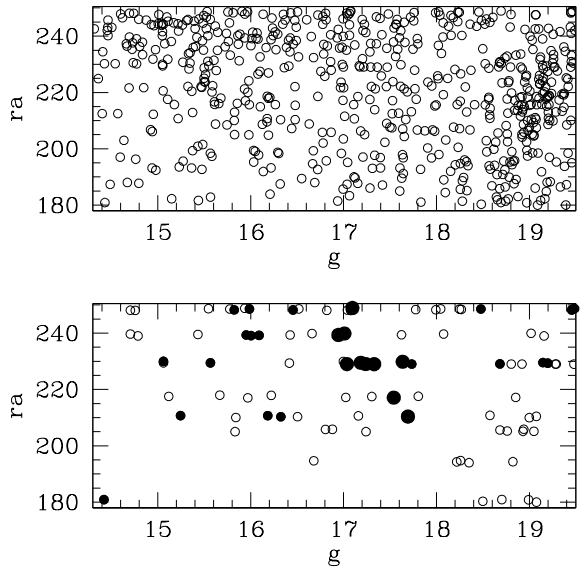}}
}
\caption{{\em Upper:} A plot of RA against $g$ for all the BHB star
candidates in the SDSS satisfying our selection criteria (detailed in
the text). Selection was confined to the equatorial strip and the
range $180^\circ<$RA$<250^\circ$ . There are 610 stars satisfying the
selection criteria. {\em Lower:} A plot of RA against $g$ for the 96
stars observed. The coordinates of the eight pointings are provided in
Table \ref{pointings}. Filled circles are stars that are classified
BHB. The large filled circles are candidate stream members.}
\label{select_ra_dec}
\end{figure}

In selecting candidate distant BHB stars from the SDSS, we used
the results of Yanny et al. (2000), who studied the spatial distribution
of a sample of A--type stars selected from the SDSS using the
(reddening--corrected) colour selection box $-0.3<g-r<0.0$,
$0.8<u-g<1.4$. On this basis, we adopted the colour cuts defined by
Yanny et al. (2000), and limited candidate selection to objects with
colour error $\sigma(g-r)<0.07$. There are 610 objects in this
sample. While these colour cuts should be nearly optimal in terms of
the fraction of candidates that are A--type stars, we still
expect substantial contamination by blue stragglers.

We use two methods to classify stars into categories BHB and blue
straggler. As one method makes use of $(B-V)_0$ colours we need to
convert the extinction corrected $g-r$ colours to $B-V$. In Clewley et
al. (2005) we described a re-investigation of this conversion and
compared it to Fukugita et al. (1996) and Smith (2002). We derived a
cubic relation that provides an improved fit. The transformation is
given by \be B-V = 0.764(g-r) - 0.170(g-r)^2 + 0.715(g-r)^3 +
0.218.\ee We stress that this relation is specifically for A--type
stars, and is not expected to be reliable for other types of star.

We chose 8 pointings roughly equidistant in RA along the stream. The
coordinates of the centres of the pointings are provided in
Table \ref{pointings}. The final selection includes 96 A--type stars
which are listed in Table \ref{tab2}. In this table Column 1 is our running
number, and column 2 lists the coordinates.  Successive columns
provide the dereddened SDSS $g$ magnitude, and the dereddened $u-g$
and $g-r$ colours.  The last column provides the dereddened $B-V$
colour, calculated using the transformation shown in Equation 1.

Looking ahead, our classification methods were developed specifically
for objects with strong Balmer lines, defined by EW H$\gamma>13$\AA,
and with a signal-to-noise (S/N) in the continuum $>15$\AA$^{-1}$. Of
the 96 selected candidates we are able to adequately classify 44 stars,
of these 29 were classified BHB.  For the 52 stars that we are unable
to classify the majority (37 stars) were because the S/N was
insufficient. In addition, there were 4 QSOs and 11 stars with EW
H$\gamma<13$\AA. 

Figure \ref{select_ra_dec} shows the distribution in RA and $g$
magnitude of the selections. In Figure \ref{select_ra_dec}(lower) the
96 colour selected stars are shown as open circles with solid circles
representing the 29 stars that are classified BHB. The horizontal stripes
represent the pointings.

\begin{table}
\centering
\begin{tabular}{cccc}
\hline\\[-12pt]
No.  & \multicolumn{1}{c}{$l$} & \multicolumn{1}{c}{$b$} & \multicolumn{1}{c}{RA}  \\
  & & &
\multicolumn{1}{r}{J 2000} \\
\hline\\[-12pt]  
01 & 277.27  & 60.38 & 12 02  \\
02 & 329.12  & 60.28 & 13 42  \\
03 & 345.38  & 55.23 & 14 22  \\
04 & 355.34  & 49.97 & 14 54  \\
05 &   1.42  & 45.57 & 15 18  \\
06 &   6.59  & 40.90 & 15 42  \\
07 &   9.64  & 37.66 & 15 58  \\
08 &  15.00  & 31.00 & 16 30  \\
\hline 
\end{tabular}
\caption{Galactic and equatorial coordinates of the centres of the
eight fields observed in the survey, ordered by Right Ascension. The
declination of all eight fields are centered along the equator.\label{pointings}}
\end{table}

\begin{table*}
\begin{center}
\begin{center}
\centering
\begin{tabular}{cccccc}
\hline
\noalign{\smallskip}   
No. & Identification (J2000) & $g$ & ($u-g$)$_0$ & ($g-r$)$_0$ & ($B-V$)$_0$ \\
(1) & (2) & (3) & (4) & (5) & (6) \\
\hline  
01 & J120341.440+001400.96   &  18.703   $\pm$  0.020  & 1.092  $\pm$  0.037  &  -0.024   $\pm$    0.026  &    0.200  $\pm$  0.0200   \\ 
02 & J155953.064+002931.70   &  19.018   $\pm$  0.015  & 1.078  $\pm$  0.054  &  -0.094   $\pm$    0.021  &    0.144  $\pm$  0.0169   \\ 
03 & J163544.242+001048.25   &  17.095   $\pm$  0.010  & 1.132  $\pm$  0.023  &  -0.192   $\pm$    0.014  &    0.060  $\pm$  0.0126   \\ 
04 & J143156.510+001251.44   &  15.668   $\pm$  0.022  & 1.105  $\pm$  0.030  &  -0.117   $\pm$    0.026  &    0.125  $\pm$  0.0214   \\ 
05 & J151955.913+001617.05   &  15.061   $\pm$  0.015  & 1.204  $\pm$  0.017  &  -0.054   $\pm$    0.020  &    0.176  $\pm$  0.0157   \\ 
06 & J120339.027+000522.44   &  14.421   $\pm$  0.046  & 1.338  $\pm$  0.060  &  -0.096   $\pm$    0.054  &    0.143  $\pm$  0.0431   \\ 
07 & J163518.461-001113.78   &  15.936   $\pm$  0.008  & 1.051  $\pm$  0.020  &  -0.224   $\pm$    0.012  &    0.031  $\pm$  0.0113   \\ 
08 & J125902.040+001331.21   &  18.259   $\pm$  0.022  & 1.236  $\pm$  0.039  &  -0.212   $\pm$    0.033  &    0.042  $\pm$  0.0301   \\ 
09 & J134332.838+001243.41   &  18.940   $\pm$  0.018  & 1.184  $\pm$  0.048  &  -0.093   $\pm$    0.025  &    0.145  $\pm$  0.0201   \\ 
\noalign{\smallskip}    						       
\hline 
\end{tabular}
\end{center}
\caption{Photometric data for the 96 target BHB candidates taken from
  the SDSS. Column (1) is our running number, column (2) lists the
  coordinates. Columns (3) to (5) list the dereddened $g$ magnitude,
  $u-g$, $g-r$ colours. The last column provides the $(B-V)_{0}$
  colour. The table is presented in its entirety in the electronic
  edition of Monthly Notices.\label{tab2}}
\end{center}
\end{table*}

\section{Spectroscopic observations}
Medium resolution spectra were obtained for the 96 candidates in eight
pointings spread over four nights from 11-14 May 2005 using the
two-degree field (2dF; Lewis et al. 2002) multi-object spectrograph on
the 3.9-m Anglo-Australian Telescope (AAT). The instrument was
equipped with a $2048^2$ Tek CCD, with a projected scale of 0.2\arcsec
pixel$^{-1}$.  We used the R1200B grating, giving a dispersion of 1.1
{\AA} pixel$^{-1}$ and a spectral coverage of 3800--4900 {\AA}, which
includes the relevant lines H$\delta$, H$\gamma$, and Ca II K
$\lambda3933$\AA.  With the 1\arcsec fibres the FWHM resolution, measured
by fitting Gaussian profiles to strong unblended arc lines, was
2.6 {\AA} which is sufficient for the line--fitting procedure.  Four
BHB radial velocity standard stars in the globular cluster M5 were
also observed nightly. Table \ref{m5_gc} summarizes relevant
information for the standards. This table provides a list of the
identification, RA and Dec., $V$ magnitude, $(B-V)_{0}$ colour, and
the heliocentric radial velocity, V$_{\odot}$. Successive columns (6)
to (11) in Table \ref{m5_gc} contain averages of $H\delta$ and
$H\gamma$ lines measured from a S\'{e}rsic function discussed in more
detail in \S4.

Total integration times between 7200 and 12000 seconds were chosen
(split into 2400 second exposures) using the 2dF exposure-time
calculator, based on the seeing, transparency, and lunar phase. This
was done in order to achieve a minimum continuum $S/N$ ratio of 15
${\mathrm\AA}^{-1}$, which was required to classify the stars (Clewley
et al. 2002). In the event, observations of 37 of the 96 targets
failed to achieve the required $S/N$, and therefore these targets
cannot be reliably classified.  The failures were primarily in the
cases where the seeing was poor and the targets were faint. All
targets were observed near culmination, with a mean airmass of
$1.2\pm0.1$.
\begin{figure}
\centering{ \scalebox{0.35}{
\includegraphics*[-50,140][700,700]{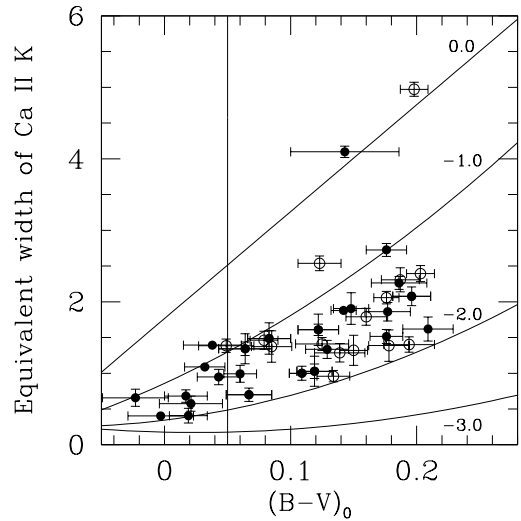}}}
\caption{CaII K line ($3933\,$\AA) EW(\AA) for the sample plotted against
$(B-V)_0$. The curves represent lines of constant metallicity for
[Fe/H] = -1.0, -2.0 and -3.0 taken from Wilhelm et al. (1999). The
straight line represents a best fit to stars in the Pleiades and Coma
clusters assumed to be of solar metallicity. The vertical line at
$(B-V)_0$ = 0.05 is the limit for which metallicities can be
determined. The 29 stars classified as BHB stars are marked by filled
circles, the 15 stars classified blue straggler are marked by open
circles.}
\label{fig_cak}
\end{figure}

The image frames were reduced with the 2dF pipeline
(2dfDR v3.31, Bailey et al. 2003). This package carries out bias
subtraction, flat-fielding, tram-line mapping to the fibre locations
on the CCD, fibre extraction, arc identification and wavelength
calibration. For the tram-line mapping and wavelength calibration,
fibre flat-field frames and CuAr arc observations were made before and
after each 45~min exposure. The arcs were used to derive accurate
positions of the spectra on the CCD and to calculate the time-varying
dispersion solution due to flexure of the instrument. This approach
was very successful. We found, from the scatter of the radial velocity
standard stars (table \ref{m5_gc}), a {\em rms} drift in the zero
point of $9\,$kms$^{-1}$ over the entire data set.

\begin{table*}   
\begin{flushleft}
\begin{center}
\begin{tabular}{lcccccccccc}
\hline
\noalign{\smallskip}   
\multicolumn{1}{c}{ID} &   
\multicolumn{1}{c}{RA (J2000) Dec.} &   
\multicolumn{1}{c}{$V_0$} &     
\multicolumn{1}{c}{$(B-V)_0$} &   
\multicolumn{1}{c}{V$_{\odot}$} &   
\multicolumn{1}{c}{$D_{0.15}(\gamma\delta)$} & 
\multicolumn{1}{c}{$b(\gamma\delta)$} &       
\multicolumn{1}{c}{$c(\gamma\delta)$} &
\multicolumn{1}{c}{$A$} &
\multicolumn{1}{c}{$B$} & 
\multicolumn{1}{c}{$\theta$} \\   
\multicolumn{1}{c}{} &  
\multicolumn{1}{c}{} &
\multicolumn{1}{c}{} &
\multicolumn{1}{c}{} &   
\multicolumn{1}{c}{[km s$^{-1}$]} &      
\multicolumn{1}{c}{[\AA]} &      
\multicolumn{1}{c}{[\AA]} &    
\multicolumn{1}{c}{} &
\multicolumn{1}{c}{} &
\multicolumn{1}{c}{} &  
\multicolumn{1}{c}{} \\    
\multicolumn{1}{c}{(1)} &  
\multicolumn{1}{c}{(2)} &
\multicolumn{1}{c}{(3)} &
\multicolumn{1}{c}{(4)} &   
\multicolumn{1}{c}{(5)} &      
\multicolumn{1}{c}{(6)} &      
\multicolumn{1}{c}{(7)} &    
\multicolumn{1}{c}{(8)} &
\multicolumn{1}{c}{(9)} &
\multicolumn{1}{c}{(10)} &
\multicolumn{1}{c}{(11)} \\
\noalign{\smallskip}   
\hline  
\noalign{\smallskip}   
M5-II-78    & J$151826.93+020717.78$ & 14.95  & 0.12  & 42.2 $\pm$ 1.1 &  30.322 $\pm$  0.501  &  7.778  &  0.799 &   0.196 &   0.021  &  1.5139  \\
	    &			     &	      &       &		       &  30.276 $\pm$  0.459  &  7.966  &  0.829 &   0.184 &   0.021  &  1.5131  \\ \hline
M5-IV-05    & J$151835.34+020227.94$ & 15.15  & 0.15  & 56.9 $\pm$ 1.2 &  31.675 $\pm$  0.482  &  8.365  &  0.836 &   0.192 &   0.020  &  1.5164  \\
	    &			     &	      &       &		       &  31.510 $\pm$  0.500  &  8.421  &  0.851 &   0.197 &   0.022  &  1.5171  \\ \hline
M5-III-69   & J$151830.43+020224.57$ & 15.06  & 0.18  & 56.2 $\pm$ 1.5 &  28.654 $\pm$  0.489  &  7.196  &  0.775 &   0.189 &   0.020  &  1.5113  \\
	    &			     &	      &       &		       &  29.282 $\pm$  0.465  &  7.248  &  0.754 &   0.179 &   0.018  &  1.5131  \\ \hline
M5-I-53     & J$151836.35+020744.60$ & 15.06  & 0.06  & 52.2 $\pm$ 1.4 &  31.966 $\pm$  0.454  &  8.372  &  0.827 &   0.182 &   0.019  &  1.5173  \\
\noalign{\smallskip}   
\hline 
\end{tabular}
\caption{Spectroscopic measurements of four M5 globular cluster BHB
  stars. The names are from Arp (1955) and Arp (1962), the photometry
  is from Cudworth (1979), and the radial velocities are from Peterson
  (1983). Columns (1) and (2) are the number and coordinates.  Columns
  (3) and (4) list the dereddened $V$ magnitude and $B-V$ colours.
  The radial velocity, corrected to the heliocentric frame, is
  provided in column (5). Successive columns (6) to
  (11) contain averages of $H\delta$ and $H\gamma$ line measurements
  measured from a S\'{e}rsic function. Column (6) is the line width at
  a depth $15\%$ below the continuum and the parameters in (7) and (8)
  are the scale width ($b$), and the shape index ($c$). The errors in
  these latter two quantities are shown in (9) to (11), these are $A$
  and $B$, the semi-major and semi-minor axes of the error ellipse in
  the $b-c$ plane, and $\theta$ the orientation of the semi-major
  axis, measured anti-clockwise from the $b$-axis.  Here the error
  corresponds to the $68\%$ confidence interval for each axis in
  isolation [see Clewley et al. (2002) for further details].
\label{m5_gc}}
 
\end{center}
\end{flushleft}
\end{table*}
\begin{table*}
\begin{center}
\begin{tiny}
\begin{tabular}{lrcrccccrcrrl}
\hline
\noalign{\smallskip}   
\multicolumn{1}{c}{No.} &   
\multicolumn{1}{c}{S/N} &   
\multicolumn{1}{c}{$D_{0.15}(\gamma\delta)$} &   
\multicolumn{1}{c}{$b(\gamma\delta)$} &    
\multicolumn{1}{c}{$c(\gamma\delta)$} &
\multicolumn{1}{c}{$A$} &  
\multicolumn{1}{c}{$B$} &  
\multicolumn{1}{c}{$\theta$} &    
\multicolumn{1}{c}{EW(CaIIK)} &   
\multicolumn{1}{c}{[Fe/H]} & 
\multicolumn{1}{c}{V$_{\odot}$} &     
\multicolumn{1}{c}{R$_{\odot}$} &   
\multicolumn{1}{c}{Class.} \\
\multicolumn{1}{c}{} &  
\multicolumn{1}{c}{[\AA]$^{-1}$} &  
\multicolumn{1}{c}{[\AA]} &
\multicolumn{1}{c}{[\AA]} &
\multicolumn{1}{c}{} &  
\multicolumn{1}{c}{} &  
\multicolumn{1}{c}{} &  
\multicolumn{1}{c}{} &
\multicolumn{1}{c}{[\AA]} &   
\multicolumn{1}{c}{} &   
\multicolumn{1}{c}{[km s$^{-1}$]} &  
\multicolumn{1}{c}{[kpc]} &  
\multicolumn{1}{c}{} \\
\multicolumn{1}{c}{(1)} &  
\multicolumn{1}{c}{(2)} &  
\multicolumn{1}{c}{(3)} &
\multicolumn{1}{c}{(4)} &
\multicolumn{1}{c}{(5)} &  
\multicolumn{1}{c}{(6)} &  
\multicolumn{1}{c}{(7)} &  
\multicolumn{1}{c}{(8)} &
\multicolumn{1}{c}{(9)} &   
\multicolumn{1}{c}{(10)} &   
\multicolumn{1}{c}{(11)} &  
\multicolumn{1}{c}{(12)} &  
\multicolumn{1}{l}{(13)} \\
\noalign{\smallskip}   
\hline  
\noalign{\smallskip} 
03  & 39.96  & 31.90  $\pm$  0.43 &   9.44  &  1.00  &  0.16  &   0.025  &   1.526   &  0.992  $\pm$  0.13  & -1.42  $\pm$ 0.16  &  -210.19  $\pm$ 09.49 & 17.16 $\pm$	1.07  & BHB    	\\
04  & 34.59  & 40.47  $\pm$  0.45 &  10.09  &  0.76  &  0.19  &   0.013  &   1.529   &  1.411  $\pm$  0.13  & -1.49  $\pm$ 0.27  &  -256.56  $\pm$ 08.20 & 4.42  $\pm$	0.58  & A/BS    \\ 	
05  & 45.28  & 24.90  $\pm$  0.33 &   6.18  &  0.74  &  0.13  &   0.015  &   1.503   &  2.726  $\pm$  0.12  & -0.99  $\pm$ 0.29  &   -41.56  $\pm$ 05.17 & 7.18  $\pm$	0.45  & BHB    	\\
06  & 32.42  & 26.57  $\pm$  0.41 &   6.36  &  0.73  &  0.16  &   0.016  &   1.508   &  4.098  $\pm$  0.13 &$\,$0.00 $\pm$ 0.51  &    60.82  $\pm$ 10.62 & 5.28  $\pm$	0.33  & BHB    	\\
12  & 23.40  & 30.43  $\pm$  0.74 &   8.33  &  0.88  &  0.28  &   0.036  &   1.517   &  1.343  $\pm$  0.14  & -1.09  $\pm$ 0.37  &  -122.20  $\pm$ 13.04 & 16.56 $\pm$	1.04  & BHB    	\\
13  & 45.55  & 37.64  $\pm$  0.40 &  10.04  &  0.85  &  0.16  &   0.015  &   1.524   &  0.959  $\pm$  0.12  & -1.89  $\pm$ 0.11  &    94.01  $\pm$ 09.00 & 8.38  $\pm$	1.13  & A/BS   	\\
14  & 33.77  & 38.67  $\pm$  0.47 &  10.14  &  0.82  &  0.19  &   0.016  &   1.524   &  1.385  $\pm$  0.13  &  ...  ...  ...     &   -74.77  $\pm$ 06.16 & 9.09  $\pm$	0.96  & A/BS   	\\
16  & 21.95  & 32.93  $\pm$  0.69 &   8.94  &  0.87  &  0.27  &   0.031  &   1.518   &  1.608  $\pm$  0.15  & -1.31  $\pm$ 0.28  &   -85.53  $\pm$ 08.73 & 22.99 $\pm$	1.44  & BHB    	\\
17  & 15.22  & 37.35  $\pm$  1.10 &  11.23  &  1.01  &  0.44  &   0.053  &   1.531   &  1.374  $\pm$  0.17  & -1.23  $\pm$ 0.31  &   -46.94  $\pm$ 17.74 & 7.61  $\pm$	0.90  & A/BS   	\\
18  & 12.51  & 25.45  $\pm$  1.28 &   5.99  &  0.67  &  0.47  &   0.050  &   1.509   &  2.668  $\pm$  0.18  & -1.13  $\pm$ 0.26  &   -38.52  $\pm$ 14.58 & 55.48 $\pm$	3.47  & BHB?    \\	
20  & 25.18  & 34.61  $\pm$  0.65 &   8.07  &  0.70  &  0.26  &   0.019  &   1.522   &  1.286  $\pm$  0.14  & -1.67  $\pm$ 0.24  &   102.15  $\pm$ 09.83 & 6.83  $\pm$	0.93  & A/BS   	\\
21  & 28.40  & 36.53  $\pm$  0.53 &   8.54  &  0.72  &  0.21  &   0.015  &   1.525   &  1.402  $\pm$  0.14  & -1.88  $\pm$ 0.21  &    73.66  $\pm$ 05.07 & 4.91  $\pm$	0.77  & A/BS   	\\
22  & 15.09  & 26.39  $\pm$  1.16 &   6.39  &  0.72  &  0.45  &   0.048  &   1.509   &  1.861  $\pm$  0.18  & -1.50  $\pm$ 0.24  &    -1.47  $\pm$ 12.58 & 47.12 $\pm$	2.95  & BHB    	\\
24  & 70.09  & 28.30  $\pm$  0.23 &   8.94  &  1.16  &  0.08  &   0.019  &   1.529   &  0.565  $\pm$  0.11  & -1.58  $\pm$ 0.07  &   -45.74  $\pm$ 09.75 & 7.80  $\pm$	0.49  & BHB?    \\	
25  & 27.50  & 34.55  $\pm$  0.60 &   8.37  &  0.74  &  0.24  &   0.019  &   1.524   &  1.790  $\pm$  0.14  & -1.44  $\pm$ 0.25  &     9.73  $\pm$ 06.20 & 6.91  $\pm$	1.00  & A/BS   	\\
26  & 30.97  & 41.41  $\pm$  0.51 &  11.19  &  0.86  &  0.20  &   0.017  &   1.530   &  2.538  $\pm$  0.13  & -0.67  $\pm$ 0.35  &    21.22  $\pm$ 07.40 & 2.32  $\pm$	0.30  & A/BS   	\\
28  & 44.21  & 32.60  $\pm$  0.45 &   9.79  &  1.05  &  0.17  &   0.029  &   1.527   &  1.494  $\pm$  0.12  & -1.46  $\pm$ 0.13  &  -150.49  $\pm$ 07.19 & 5.98  $\pm$	0.37  & BHB?    \\	
29  & 36.83  & 31.44  $\pm$  0.47 &   7.41  &  0.72  &  0.18  &   0.015  &   1.521   &  2.019  $\pm$  0.13  & -1.60  $\pm$ 0.11  &    34.95  $\pm$ 04.89 & 5.43  $\pm$	0.89  & A/BS?   \\
30  & 50.47  & 31.73  $\pm$  0.28 &   9.09  &  0.96  &  0.11  &   0.015  &   1.515   &  0.575  $\pm$  0.12  &  ...  ...  ...     &    84.78  $\pm$ 08.22 & 6.99  $\pm$	0.44  & BHB    	\\
31  & 24.74  & 33.37  $\pm$  0.59 &   9.61  &  0.97  &  0.23  &   0.033  &   1.516   &  0.948  $\pm$  0.14  & -1.33  $\pm$ 0.36  &   -31.73  $\pm$ 07.07 & 17.57 $\pm$	1.10  & BHB    	\\
32  & 36.99  & 31.38  $\pm$  0.52 &   9.72  &  1.12  &  0.18  &   0.035  &   1.533   &  1.104  $\pm$  0.13  & -1.47  $\pm$ 0.14  &    27.57  $\pm$ 08.63 & 8.14  $\pm$	0.51  & BHB?    \\	
33  & 17.23  & 44.10  $\pm$  1.05 &  11.98  &  0.88  &  0.41  &   0.035  &   1.533   &  1.321  $\pm$  0.16  & -1.71  $\pm$ 0.12  &  -236.21  $\pm$ 12.42 & 12.51 $\pm$	1.76  & A/BS   	\\
35  & 30.54  & 29.52  $\pm$  0.44 &   8.53  &  0.98  &  0.17  &   0.027  &   1.512   &  0.657  $\pm$  0.13  &  ...  ...  ...     &  -124.38  $\pm$ 09.08 & 9.94  $\pm$	0.62  & BHB    	\\
36  & 70.46  & 29.53  $\pm$  0.21 &   8.57  &  0.97  &  0.08  &   0.013  &   1.510   &  0.407  $\pm$  0.11  &  ...  ...  ...     &   -79.52  $\pm$ 10.22 & 8.08  $\pm$	0.51  & BHB    	\\
42  & 34.81  & 33.70  $\pm$  0.49 &   7.97  &  0.72  &  0.19  &   0.014  &   1.522   &  4.973  $\pm$  0.13 &$\,$0.00  $\pm$ 0.00 &    18.81  $\pm$ 14.56 & 2.85  $\pm$	0.45  & A/BS   	\\
43  & 17.93  & 31.68  $\pm$  1.08 &   7.96  &  0.78  &  0.44  &   0.039  &   1.517   &  1.028  $\pm$  0.16  & -1.77  $\pm$ 0.10  &   160.75  $\pm$ 15.77 & 16.69 $\pm$	1.04  & BHB    	\\
44  & 19.96  & 28.90  $\pm$  0.85 &   6.88  &  0.72  &  0.33  &   0.030  &   1.513   &  1.515  $\pm$  0.15  & -1.72  $\pm$ 0.11  &   104.07  $\pm$ 10.50 & 34.63 $\pm$	2.17  & BHB    	\\
47  & 39.99  & 37.10  $\pm$  0.43 &   8.90  &  0.73  &  0.17  &   0.012  &   1.525   &  2.058  $\pm$  0.13  & -1.37  $\pm$ 0.12  &    18.68  $\pm$ 06.50 & 5.59  $\pm$	0.84  & A/BS   	\\
48  & 32.27  & 29.68  $\pm$  0.44 &   8.58  &  0.98  &  0.17  &   0.027  &   1.509   &  0.681  $\pm$  0.13  & ...  ...  ...      &   191.31  $\pm$ 09.09 & 9.62  $\pm$	0.60  & BHB    	\\
49  & 71.92  & 32.40  $\pm$  0.25 &   9.95  &  1.09  &  0.10  &   0.018  &   1.530   &  0.653  $\pm$  0.11  & -1.84  $\pm$ 0.11  &  -179.80  $\pm$ 15.74 & 9.44  $\pm$	0.59  & BHB?    \\	
51  & 10.34  & 35.38  $\pm$  1.68 &   8.46  &  0.72  &  0.69  &   0.050  &   1.520   &  1.447  $\pm$  0.20  & -1.68  $\pm$ 0.23  &    -9.55  $\pm$ 18.51 & 8.12  $\pm$	1.17  & A/BS?   \\
58  & 16.50  & 26.50  $\pm$  0.89 &   6.51  &  0.71  &  0.35  &   0.038  &   1.504   &  1.335  $\pm$  0.16  & -1.57  $\pm$ 0.25  &    70.13  $\pm$ 14.54 & 37.35 $\pm$	2.34  & BHB    	\\
63  & 12.25  & 37.52  $\pm$  1.42 &   9.23  &  0.72  &  0.55  &   0.046  &   1.526   &  0.506  $\pm$  0.18  & -2.93  $\pm$ 0.08  &  -184.24  $\pm$ 21.74 & 12.19 $\pm$	1.99  & A/BS?   \\
64  & 27.34  & 23.89  $\pm$  0.60 &   5.22  &  0.62  &  0.30  &   0.049  &   1.505   &  2.034  $\pm$  0.14  & -1.17  $\pm$ 0.43  &    38.66  $\pm$ 12.05 & 19.34 $\pm$	1.21  & BHB?    \\	
66  & 40.67  & 25.95  $\pm$  0.38 &   5.81  &  0.67  &  0.14  &   0.013  &   1.510   &  1.619  $\pm$  0.12  & -1.83  $\pm$ 0.20  &   -16.48  $\pm$ 06.52 & 7.47  $\pm$	0.47  & BHB    	\\
67  & 34.66  & 28.71  $\pm$  0.44 &   7.22  &  0.78  &  0.18  &   0.018  &   1.508   &  1.905  $\pm$  0.13  & -1.28  $\pm$ 0.13  &  -214.94  $\pm$ 06.28 & 11.41 $\pm$	0.71  & BHB    	\\
68  & 27.66  & 41.95  $\pm$  0.63 &  11.21  &  0.85  &  0.25  &   0.021  &   1.529   &  1.462  $\pm$  0.14  & -1.11  $\pm$ 0.17  &  -218.00  $\pm$ 08.86 & 12.89 $\pm$	1.50  & A/BS   	\\
70  & 32.80  & 30.96  $\pm$  0.43 &   8.69  &  0.93  &  0.17  &   0.023  &   1.514   &  1.486  $\pm$  0.13  & -1.12  $\pm$ 0.34  &   -95.59  $\pm$ 06.81 & 18.70 $\pm$	1.17  & BHB    	\\
72  & 15.00  & 25.67  $\pm$  1.09 &   6.27  &  0.68  &  0.45  &   0.045  &   1.502   &  2.263  $\pm$  0.18  & -1.31  $\pm$ 0.25  &    49.97  $\pm$ 17.98 & 24.22 $\pm$	1.52  & BHB    	\\
73  & 15.53  & 26.14  $\pm$  0.99 &   8.14  &  1.13  &  0.41  &   0.089  &   1.494   &  0.374  $\pm$  0.19  & -1.71  $\pm$ 0.44  &   -10.93  $\pm$ 17.91 & 17.24 $\pm$	1.08  & BHB    	\\
74  & 12.07  & 25.67  $\pm$  1.30 &   5.53  &  0.58  &  0.50  &   0.036  &   1.511   &  2.659  $\pm$  0.18  & -1.04  $\pm$ 0.28  &   100.23  $\pm$ 15.48 & 50.10 $\pm$	3.14  & BHB?    \\	
75  & 15.93  & 32.21  $\pm$  0.98 &   8.83  &  0.88  &  0.38  &   0.044  &   1.521   &  1.171  $\pm$  0.17  & -2.06  $\pm$ 0.15  &   -11.71  $\pm$ 15.06 & 11.14 $\pm$	0.70  & BHB    	\\
76  & 20.07  & 33.22  $\pm$  0.84 &   7.79  &  0.71  &  0.33  &   0.024  &   1.523   &  2.308  $\pm$  0.15  & -1.29  $\pm$ 0.12  &    53.50  $\pm$ 07.38 & 11.27 $\pm$	1.74  & A/BS   	\\
77  & 11.21  & 41.70  $\pm$  1.64 &  11.80  &  0.93  &  0.72  &   0.066  &   1.520   &  1.346  $\pm$  0.19  & -1.71  $\pm$ 0.23  &    -1.98  $\pm$ 16.19 & 18.60 $\pm$	2.63  & A/BS?   \\
78  & 54.50  & 31.09  $\pm$  0.31 &   8.75  &  0.93  &  0.12  &   0.017  &   1.519   &  1.879  $\pm$  0.12  & -1.26  $\pm$ 0.14  &    64.57  $\pm$ 03.66 & 10.06 $\pm$	0.63  & BHB    	\\
79  & 16.84  & 35.34  $\pm$  1.10 &   8.73  &  0.76  &  0.46  &   0.035  &   1.523   &  1.387  $\pm$  0.16  & -1.82  $\pm$ 0.20  &   -17.16  $\pm$ 18.60 & 5.79  $\pm$	0.87  & A/BS   	\\
80  & 13.43  & 29.85  $\pm$  1.20 &   6.57  &  0.61  &  0.48  &   0.040  &   1.516   &  2.778  $\pm$  0.17  & -0.74  $\pm$ 0.34  &   -13.19  $\pm$ 20.07 & 39.86 $\pm$	2.49  & BHB?    \\
81  & 48.89  & 29.55  $\pm$  0.34 &   8.59  &  0.98  &  0.12  &   0.021  &   1.519   &  1.000  $\pm$  0.12  & -1.74  $\pm$ 0.11  &    84.43  $\pm$ 07.77 & 13.26 $\pm$	0.83  & BHB    	\\
82  & 22.99  & 31.67  $\pm$  0.65 &   8.92  &  0.92  &  0.26  &   0.034  &   1.516   &  1.393  $\pm$  0.14  &  ...  ...  ...     &    40.56  $\pm$ 08.55 & 11.75 $\pm$	0.74  & BHB    	\\
83  & 26.32  & 27.84  $\pm$  0.62 &   7.16  &  0.79  &  0.23  &   0.028  &   1.513   &  2.077  $\pm$  0.14  & -1.48  $\pm$ 0.12  &   -25.56  $\pm$ 05.57 & 17.97 $\pm$	1.12  & BHB    	\\
87  & 12.54  & 26.92  $\pm$  1.33 &   5.67  &  0.61  &  0.51  &   0.034  &   1.515   &  1.618  $\pm$  0.18  & -1.71  $\pm$ 0.22  &   -66.15  $\pm$ 17.24 & 50.53 $\pm$	3.16  & BHB?    \\
88  & 39.76  & 28.17  $\pm$  0.41 &   6.29  &  0.66  &  0.15  &   0.012  &   1.515   &  1.588  $\pm$  0.13  & -1.88  $\pm$ 0.10  &  -114.59  $\pm$ 07.40 & 6.81  $\pm$	1.14  & A/BS?   \\
92  & 95.40  & 27.89  $\pm$  0.17 &   8.56  &  1.09  &  0.06  &   0.013  &   1.516   &  0.595  $\pm$  0.11  & -1.52  $\pm$ 0.06  &   -48.73  $\pm$ 10.09 & 5.27  $\pm$	0.33  & BHB?    \\	
93  & 32.73  & 32.11  $\pm$  0.43 &   8.80  &  0.88  &  0.17  &   0.020  &   1.514   &  1.088  $\pm$  0.13  & ...  ...  ...      &  -147.60  $\pm$ 09.00 & 9.33  $\pm$	0.58  & BHB    	\\
94  & 15.04  & 29.84  $\pm$  1.22 &   8.65  &  0.98  &  0.44  &   0.074  &   1.515   &  0.403  $\pm$  0.18  & ...  ...  ...      &    13.16  $\pm$ 22.97 & 19.33 $\pm$	1.21  & BHB    	\\
96  & 31.72  & 30.63  $\pm$  0.56 &   7.12  &  0.70  &  0.22  &   0.017  &   1.516   &  2.394  $\pm$  0.13  & -1.34  $\pm$ 0.12  &   -51.85  $\pm$ 06.27 & 6.34  $\pm$	1.02  & A/BS   	\\
99  & 36.32  & 30.68  $\pm$  0.40 &   8.74  &  0.95  &  0.15  &   0.023  &   1.513   &  0.700  $\pm$  0.13  & -1.78  $\pm$ 0.20  &   -56.20  $\pm$ 06.98 & 19.24 $\pm$	1.20  & BHB    	\\
100 & 94.98  & 30.73  $\pm$  0.17 &   9.62  &  1.13  &  0.06  &   0.012  &   1.526   &  0.717  $\pm$  0.11  & -1.57  $\pm$ 0.16  &  -113.25  $\pm$ 08.80 & 5.68  $\pm$	0.36  & BHB     \\
102 & 50.83  & 33.12  $\pm$  0.42 &   9.67  &  0.99  &  0.15  &   0.023  &   1.528   &  0.986  $\pm$  0.12  & -1.58  $\pm$ 0.14  &    -5.25  $\pm$ 08.00 & 17.35 $\pm$	1.09  & BHB     \\
\hline 
\end{tabular}
\caption{Spectroscopic data for the BHB star
candidates. The columns (3) to (8) and (11) are the same as in Table
\ref{m5_gc}. In addition, columns (1) and (2) give the number of the
star and the spectrum continuum $S/N$ per \AA. Column (9) is EW of the
CaII K line and column (10) is the measured metallicity for each star
with its error. The heliocentric distance is shown in column
(12). Finally, the classification of the star is provided in column (13).\label{spec_tabs}}
\end{tiny}
\end{center}
\end{table*}
\begin{figure*}
\begin{minipage}{133mm}
\epsfig{figure=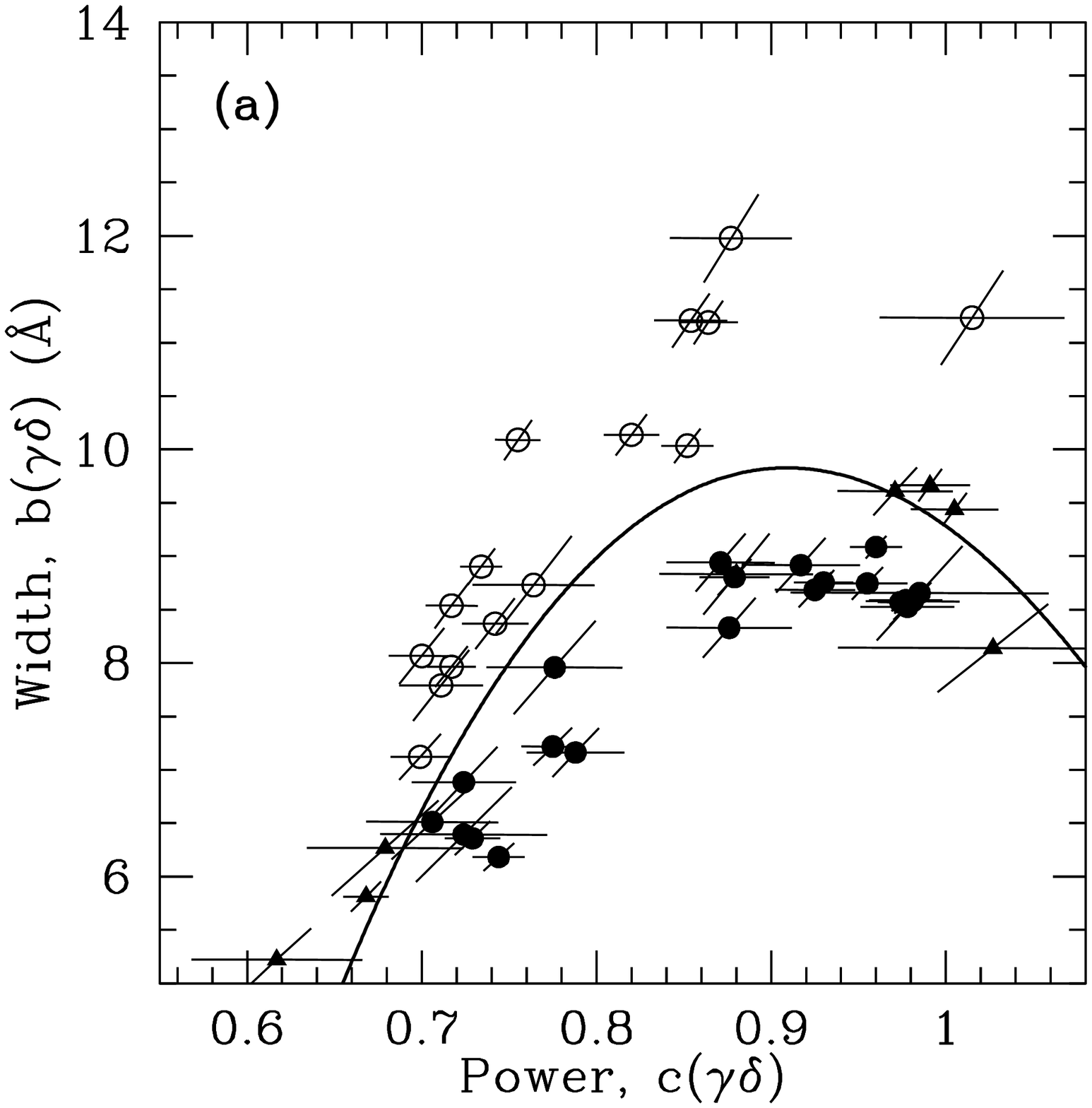,height=66mm,width=66mm}
\epsfig{figure=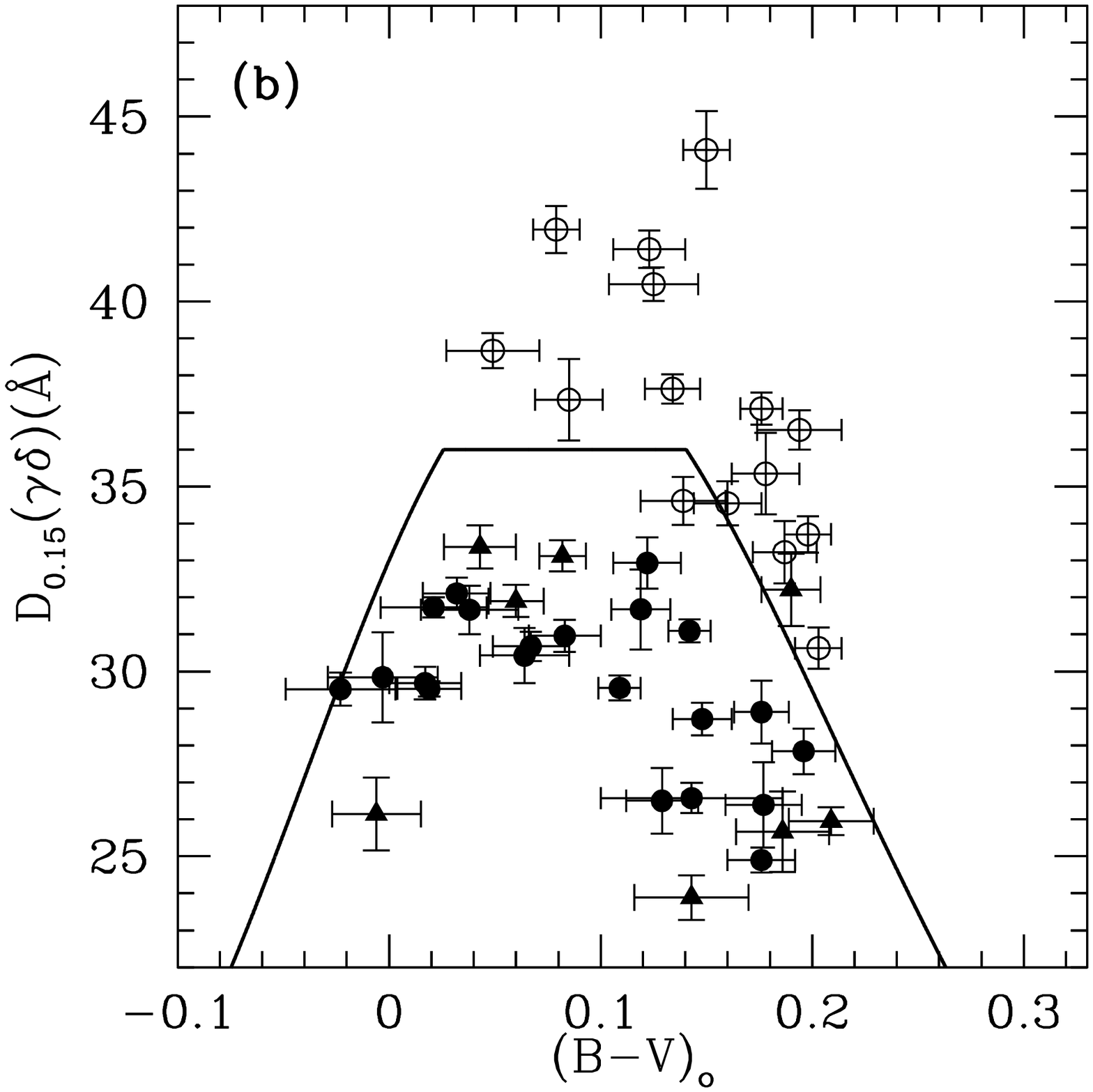,height=66mm,width=66mm}
\end{minipage}
\caption{Classification of all 44 stars with unambiguous
classifications shown in Table \ref{spec_tabs}, using the (a) the {\em
Scale width--Shape} and (b) {\em $D_{0.15}$--Colour}
classification methods. The solid curves are the classification
boundaries explained in the text. Filled circles are stars classified
BHB in both plots, i.e. the stars below the classification boundary in
each plot. Open circles are stars classified A/BS. Filled triangles
are stars below the classification boundary in only one plot but are
nonetheless classified BHB. In plot (a) there are 23 stars below the
boundary and 21 above it. In plot (b) there are 29 stars below the
boundary and 15 above it. A total of 29 stars are classified BHB.
\label{sw_cw}}
\end{figure*}

\section{Spectroscopic analysis and classification}

\subsection{Analysis}
The spectra were used, in combination with the photometry, to classify
the stars, measure the metallicity and calculate the radial
velocities. For these measurements we followed the procedures set out
in Clewley et al. (2002, 2004) exactly, and we refer the reader to
those papers for full details. Below we briefly summarize the
procedures we undertake to measure the Balmer lines, metallicities and
distances.

For the Balmer lines we normalized each spectrum to the continuum, and
fitted a S\'{e}rsic function, convolved with a Gaussian that has the
FWHM of the instrumental resolution. We have two classification
procedures. First, the {\em Scale width--Shape} method, plots the
scale width $b$, and the shape index $c$. Second, the
{\em$D_{0.15}$--Colour} method plots $D_{0.15}$, which is the line
width at a depth $15\%$ below the continuum against $(B-V)_0$.

With some exceptions (including the A metallic (Am) and peculiar (Ap)
stars) the strength of the Ca II K line at constant temperature can be
used as an indicator of the metallicity of A-type stars (e.g. Pier
1983, Beers et al. 1992, Kinman et al. 1994). The
metallicities were determined from the Ca II K 3933\AA\,, line by
minimum$-\chi^2$ fitting a Gaussian to the continuum divided spectrum
over the wavelength range 3919--3949 {\AA}. The metallicities were
derived by plotting Ca II K line EW against $(B-V)_0$ and
interpolating between lines of constant metallicity (see Figure
\ref{fig_cak}). The uncertainty is established from the uncertainties
of the two quantities plotted, and an additional uncertainty of 0.3dex
is added in quadrature. This systematic error was established by using
our methods on high $S/N$ stars with known metallicities. As the lines
in Figure \ref{fig_cak} converge towards the bluer colours we consider
this plot only reliable for colours redder than $(B-V)_0 > 0.05$. No
attempt has been made to remove the possible contribution of
interstellar Ca II K absorption from the stellar K measurements.

The absolute magnitudes of BHB stars, $M_V(BHB)$, depend on both
metallicity and colour (i.e. temperature).  Preston, Shectman \& Beers
(1991) provided an empirical Luminosity-Colour relation for BHB stars
derived from a fit to globular cluster BHB stars. Recently, this
empirical relation was re-investigated by Brown et al. (2005), who
provided a physical basis for this relationship based on theoretical
modelling. In Clewley et al. (2004) we derived a relation for the
absolute magnitude of BHBs in two steps. First, we used the results
from the Clementini et al. (2003) to determine the slope of the
relation, $M_V(RR) = \alpha + \beta[\rm{Fe/H}]$. We fixed the zero
point of the relation using the measurement by Gould \& Popowski
(1998) of the absolute magnitude of RR Lyrae stars, $M_V(RR)$ = 0.77
$\pm$ 0.13 mag at [Fe/H] = -1.60, derived from statistical parallaxes
of {\it Hipparcos} observations. Combining these two
measurements resulted in $M_V(RR)=1.112+0.214[\rm{Fe/H}]$. Second we
adopted the cubic expression determined by Preston et al. (1991) for the $(B-V)_0$ colour dependence of the difference in
absolute magnitudes between BHB and RR Lyrae stars. This produced the
final expression for the absolute magnitude of BHB stars:
\begin{eqnarray}  
M_V(BHB) &=& 1.552+0.214[\mathrm{Fe/H}]-4.423(B-V)_0 \nonumber\\ & & +
17.74(B-V)^2_0-35.73(B-V)^3_0.
\label{ch4:abs_mag}
\end{eqnarray}
Distances and associated errors are then determined using the apparent
magnitudes $V_0$, and the corresponding photometric and metallicity
errors.  To compute $V$, we used the relation
$V=g^\prime-0.53(g^{\prime}-r^{\prime})$ (Fukugita et al. 1996), here
disregarding the subtle differences between the different SDSS
magnitudes ($g, g^{\prime}, g^*$, etc.).  The result produces distance
errors of $6-10\%$ for our confirmed BHB stars. We caution that the
$M_V$--metallicity relation is still controversial and so distance
measurements may be systematically in error. We refer the interested
reader to a recent review of this subject by Cacciari \& Clementini
(2003) and Alves (2004).

The absolute magnitudes of blue stragglers have been less well
studied. As we do not analyse them here we simply
adopt the relation, $M_V(BS)=1.32+4.05(B-V)_0-0.45[\mathrm{Fe/H}]$,
derived by Kinman et al. (1994).

\subsection{Classification}
In Figure \ref{sw_cw} we plot the two diagnostic diagrams for the 44
classifiable stars in the survey. Figure \ref{sw_cw}(a)
shows the {\em Scale width--Shape} method. The line--profile
quantities $b$ and $c$, averaged for H$\gamma$ and H$\delta$ are
plotted. Figure \ref{sw_cw}(b) shows the {\em $D_{0.15}-$colour}
method. Plotted are values of $D_{0.15}$ for H$\gamma$ and H$\delta$
against $(B-V)_0$ for all the stars.  The solid lines show the
classification boundaries, taken from Clewley et al. (2002, 2004),
with high--surface gravity stars (i.e.  main--sequence A stars or blue
stragglers, hereafter A/BS) above the line, and low--surface gravity
stars (i.e. BHB stars) below the line. In both plots stars classified
BHB are plotted as solid symbols and stars classified A/BS are plotted
open. As we discuss below, the eight triangles are stars that are
classified as BHB by one classification method and not the other but
were ultimately classified as BHBs.

Figure \ref{sw_cw} shows that of the 44 candidates, 29 are classified
BHB by the {\em $D_{0.15}-$colour} method, and 23 are classified BHB
by the {\em Scale width--Shape} method. There are 21 stars classified
BHB by both methods. A total of ten stars are classified BHB by one or
other of the methods. There is clearly close agreement between the two
classification methods. For the ambiguous classifications we combine
the information provided by all the parameters. The uncertainties on
each parameter define the 2D probability distribution functions for
any point.  By integrating these functions below the classification
boundary we can compute a probability $P(BHB)$ that any star is
BHB. Of the nine stars with ambiguous classification eight are
classified BHB. We also classify the remaining 15 stars with
inadequate spectroscopic $S/N$, but for these the classifications are
given as BHB?  or A/BS? to indicate that they are not reliable. We
note that the four radial velocity standards (Table 2), previously
classified BHB from high--resolution spectroscopy, are all
unambiguously classified BHB in both plots.

\section{Results}
Of the 96 selected candidates only 44 have suitable spectroscopic
$S/N$ and H$\gamma$ EW for reliable classification. The classification
of these 44 candidates results in 29 BHB stars. We have nevertheless
followed the classification procedures for the remaining 52 objects,
but for clarity have omitted them from Figures \ref{fig_cak} and
\ref{sw_cw}. For these objects the final classifications are flagged
as questionable. The results of the measurements for all the candidate
BHB stars are provided in Table \ref{spec_tabs}.

Table \ref{data_sum} contains a summary of the kinematic properties of
the final sample of 29 reliably classified BHB stars.  Listed are the
identification, RA and Dec., heliocentric velocity and distance,
Galactic coordinates $l$ and $b$ and the Galactocentric distance and
radial velocity R$_{gal}$ and V$_{gal}$ respectively. To convert the
heliocentric quantities to Galactocentric quantities, the heliocentric
radial velocities are first corrected for solar motion by assuming a
solar peculiar velocity of ($U,V,W$) = (-9,12,7), where $U$ is
directed outward from the Galactic Centre, $V$ is positive in the
direction of Galactic rotation at the position of the Sun, and $W$ is
positive toward the North Galactic Pole.  We have assumed a circular
speed of 220 km s$^{-1}$ at the Galactocentric radius of the Sun
($R_{gal} = 8.0\,$kpc).
\begin{table*} 
\begin{center}
\begin{tiny}
\begin{tabular}{lrrrrrrrr}
\hline
\noalign{\smallskip} 
\multicolumn{1}{c}{No.} & 
\multicolumn{1}{c}{RA (J2000)} &  
\multicolumn{1}{c}{Dec.} &  
\multicolumn{1}{c}{V$_{\odot}$} &  
\multicolumn{1}{c}{R$_{\odot}$} & 
\multicolumn{1}{c}{$l$} &  
\multicolumn{1}{c}{$b$} & 
\multicolumn{1}{c}{R$_{gal}$} &
\multicolumn{1}{c}{V$_{gal}$} \\
\multicolumn{1}{c}{} & 
\multicolumn{1}{c}{[$^{\circ}$]} &  
\multicolumn{1}{c}{[$^{\circ}$]} & 
\multicolumn{1}{c}{[km s$^{-1}$]} & 
\multicolumn{1}{c}{[kpc]} &
\multicolumn{1}{c}{[$^{\circ}$]} &  
\multicolumn{1}{c}{[$^{\circ}$]} & 
\multicolumn{1}{c}{[kpc]} &
\multicolumn{1}{c}{[km s$^{-1}$]} \\
\multicolumn{1}{c}{(1)} &  
\multicolumn{1}{c}{(2)} &  
\multicolumn{1}{c}{(3)} &
\multicolumn{1}{c}{(4)} &
\multicolumn{1}{c}{(5)} &  
\multicolumn{1}{c}{(6)} &  
\multicolumn{1}{c}{(7)} &  
\multicolumn{1}{c}{(8)} &
\multicolumn{1}{c}{(9)} \\   
\noalign{\smallskip} 
\hline 
\noalign{\smallskip} 
03   & 248.9343414  &  0.1800683  &  -210.19 $\pm$  09.49  &   17.16    $\pm$  1.07  &   16.0580  &  29.8770  &  11.39  &  -143.55  \\
05   & 229.9829712  &  0.2714026  &   -41.56 $\pm$  05.17  &   7.18     $\pm$  0.45  &    2.1616  &  45.3792  &   5.91  &   -24.11  \\
06   & 180.9126129  &  0.0895671  &    60.82 $\pm$  10.62  &   5.28     $\pm$  0.33  &  277.9714  &  60.6244  &   9.28  &   -45.17  \\
12   & 239.8858643  & -0.1522919  &  -122.20 $\pm$  13.04  &   16.56    $\pm$  1.04  &    9.7586  &  37.2616  &  11.42  &   -79.61  \\
16   & 229.8426208  & -0.3234651  &   -85.53 $\pm$  08.73  &   22.99    $\pm$  1.44  &    1.3890  &  45.1006  &  18.25  &   -70.26  \\
22   & 229.5568085  &  0.1586571  &    -1.47 $\pm$  12.58  &   47.12    $\pm$  2.95  &    1.6456  &  45.6312  &  41.92  &    14.49  \\
30   & 210.7175751  &  0.3365301  &    84.78 $\pm$  08.22  &   6.99     $\pm$  0.44  &  338.5038  &  58.1867  &   7.62  &    50.33  \\
31   & 229.4858856  &  0.4894420  &   -31.73 $\pm$  07.07  &   17.57    $\pm$  1.10  &    1.9410  &  45.8996  &  13.31  &   -14.97  \\
35   & 210.7082367  &  0.0929688  &  -124.38 $\pm$  09.08  &   9.94     $\pm$  0.62  &  338.2509  &  57.9829  &   9.19  &  -159.59  \\
36   & 229.4539948  & -0.1177904  &   -79.52 $\pm$  10.22  &   8.08     $\pm$  0.51  &    1.2502  &  45.5293  &   6.22  &   -64.68  \\
43   & 239.3611908  & -0.3381351  &   160.75 $\pm$  15.77  &   16.69    $\pm$  1.04  &    9.1860  &  37.5826  &  11.56  &   201.41  \\
44   & 248.5698090  & -0.1052531  &   104.07 $\pm$  10.50  &   34.63    $\pm$  2.17  &   15.5566  &  30.0370  &  28.30  &   168.95  \\
48   & 239.3440247  &  0.3121248  &   191.31 $\pm$  09.09  &   9.62     $\pm$  0.60  &    9.8454  &  37.9719  &   6.08  &   233.88  \\
58   & 229.0891724  &  0.1762543  &    70.13 $\pm$  14.54  &   37.35    $\pm$  2.34  &    1.2279  &  45.9976  &  32.28  &    84.87  \\
66   & 217.5550232  &  0.1511529  &   -16.48 $\pm$  06.52  &   7.47     $\pm$  0.47  &  348.3371  &  54.0754  &   6.07  &   -33.16  \\
67   & 239.2290192  & -0.4766546  &  -214.94 $\pm$  06.28  &   11.41    $\pm$  0.71  &    8.9468  &  37.6097  &   7.16  &  -175.04  \\
70   & 229.0451965  & -0.1293136  &   -95.59 $\pm$  06.81  &   18.70    $\pm$  1.17  &    0.8546  &  45.8307  &  14.33  &   -81.89  \\
72   & 210.3879852  &  0.3614573  &    49.97 $\pm$  17.98  &   24.22    $\pm$  1.52  &  337.9907  &  58.3777  &  21.50  &    14.72  \\
73   & 217.5059052  &  0.3670563  &   -10.93 $\pm$  17.91  &   17.24    $\pm$  1.08  &  348.5063  &  54.2734  &  12.11  &   -27.09  \\
75   & 239.1218567  &  0.0064421  &   -11.71 $\pm$  15.06  &   11.14    $\pm$  0.70  &    9.3660  &  37.9773  &   6.28  &    29.36  \\
78   & 248.2692719  & -0.3074609  &    64.57 $\pm$  03.66  &   10.06    $\pm$  0.63  &   15.1744  &  30.1850  &   5.56  &   128.09  \\
81   & 248.2459106  & -0.3497654  &    84.43 $\pm$  07.77  &   13.26    $\pm$  0.83  &   15.1186  &  30.1823  &  11.66  &   147.76  \\
82   & 210.3329315  & -0.0795943  &    40.56 $\pm$  08.55  &   11.75    $\pm$  0.74  &  337.4755  &  58.0257  &   8.43  &     3.84  \\
83   & 229.0241089  & -0.1868408  &   -25.56 $\pm$  05.57  &   17.97    $\pm$  1.12  &    0.7725  &  45.8088  &  12.12  &   -12.09  \\
93   & 210.0473328  &  0.0858782  &  -147.60 $\pm$  09.00  &   9.33     $\pm$  0.58  &  337.1651  &  58.3123  &   6.88  &  -184.58  \\
94   & 217.1094971  &  0.4875866  &    13.16 $\pm$  22.97  &   19.33    $\pm$  1.21  &  348.1101  &  54.6178  &  16.20  &    -3.71  \\
99   & 229.0039825  & -0.2001127  &   -56.20 $\pm$  06.98  &   19.24    $\pm$  1.20  &    0.7391  &  45.8152  &  14.82  &   -42.82  \\
100  & 248.0934448  & -0.2781225  &  -113.25 $\pm$  08.80  &   5.68     $\pm$  0.36  &   15.0951  &  30.3495  &   4.53  &   -50.08  \\
102  & 248.0232239  & -0.1998732  &    -5.25 $\pm$  08.00  &   17.35    $\pm$  1.09  &   15.1287  &  30.4506  &  11.58  &    57.99  \\
\noalign{\smallskip} 
\hline 
\end{tabular}
\caption{Summary of positional and kinematic information for the BHB stars. Listed are the: identification, RA and Dec., heliocentric velocity and distance, Galactic coordinates $l$ and $b$,  and the Galactocentric distance and radial velocity R$_{gal}$ and V$_{gal}$ respectively. \label{data_sum}}
\end{tiny}
\end{center}
\end{table*}

\begin{figure}
\centering{
\scalebox{0.35}{
\includegraphics*{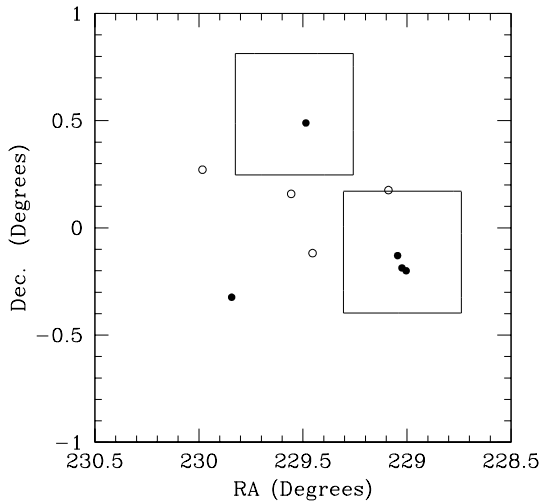}}
}
\caption{Plot of the BHB stars with similar RA and DEC. as the Pal 5
globular cluster. The two squares are regions of Pal 5 taken from Koch
et al. (2004) denoting the centre of Pal 5 (229.0,-0.1) and the
trailing stream (229.5,0.5). The filled circles are stars at a similar
distance and are therefore most likely to be associated with the
globular cluster.}
\label{glob_m5}
\end{figure}

\subsection{Comparisons with previous observations of Palomar 5}
Our survey passes over the globular cluster Palomar (Pal) 5 (RA=
229.022083$^{\circ}$, Dec.=$-$0.111389$^{\circ}$ [J2000]). This system
has a Heliocentric radial velocity, V$_{\odot}$ = $-$58.7 $\pm$ 0.20
km s$^{-1}$ (Odenkirchen et al. 2002), a distance $R_{\odot}$ =23.2
kpc and [Fe/H] = $-$1.41 (Harris 1996; Clement at al. 2001). Pal 5 is
remarkable as observations from the SDSS show a pair of tidal tails,
which extend around 4 kpc in opposite directions from the cluster and
contain more stars than the cluster itself (Odenkirchen et al. 2003;
Koch et al. 2004).

We compare our measurements of BHB stars around Pal 5 to previous work
in the literature. There are four stars (numbers 99,83,70,31; shown as
filled circles within the boxes in Figure \ref{glob_m5}) that are
plausibly associated with Pal 5, i.e. they are located either within
the projected central region of the cluster or in one of the tidal
tails. For these four stars we measure a mean heliocentric velocity of
$-$52.3 $\pm$ 31.8 and a mean metallicity of $-$1.43 $\pm$ 0.27. These
values are entirely consistent, albeit with large errors, with the
measures by Harris (1996) and Odenkirchen et al. (2003). However the
mean distances (18.4 $\pm$ 0.7 kpc) are inconsistent with those
reported in Harris (1996). Three of the four stars (numbers 99,83,70)
are located in the central position (see Figure \ref{glob_m5}) at
distances 19.2, 18.0, 18.7 respectively (mean = 18.6 $\pm$ 0.6). The
remaining star that could plausibly be in the stream (number 31) is at
$17.57 \pm 1.1$ kpc. We investigate this apparent distance discrepancy
further.
\begin{table*}
\begin{center}
\begin{center}
\centering
\begin{tabular}{ccccccc}
\hline
\noalign{\smallskip}   
ID & R.A. (J2000) & Dec. & $V$  & $B-V$ & R$_{\odot}$ & Refs. \\
\hline  
399 &  15 15 57.25 & -00 06 52.8  &  17.44 &  0.03 & 21.16 & (1), (2)  \\            
400 &  15 15 57.97 & -00 11 23.1  &  17.33 &  0.01 & 20.11 & (1)       \\                      
401 &  15 15 58.31 & -00 05 47.3  &  17.58 &  0.03 & 22.57 & (1)       \\                     
402 &  15 16 05.79 & -00 11 12.3  &  17.46 &  0.04 & 21.36 & (1)       \\     
404 &  15 16 12.79 & -00 10 02.7  &  17.38 &  0.03 & 20.58 & (1)       \\                     
\noalign{\smallskip}
\hline 
\end{tabular}
\end{center}
\caption{Summary of position, colour, magnitude and distance of RR
Lyrae stars observed in the QUEST survey (Vivas et
al. 2004). References: (1) Kinman \& Rosino (1962); (2) Clement et
al. (2001); (2) Wu et al. (2005) \label{rr_lyrae_quest}}
\end{center}
\end{table*}

\begin{table}
\centering
\begin{tabular}{ccccccc}
\hline
\noalign{\smallskip}   
ID & $V$  & $B-V$  & R$_{\odot}$ \\
\hline    
31  &  17.185 $\pm$  0.014 &  0.043 $\pm$ 0.0166 & 17.57\\  
70  &  17.238 $\pm$  0.016 &  0.083 $\pm$ 0.0174 & 18.70\\
83  &  17.036 $\pm$  0.016 &  0.196 $\pm$ 0.0147 & 17.97\\
99  &  17.329 $\pm$  0.016 &  0.067 $\pm$ 0.0178 & 19.24\\
\hline
S41 &  17.453 $\pm$  0.003 &  0.103 $\pm$ 0.0031 & 20.02\\
S40 &  17.519 $\pm$  0.002 &  0.087 $\pm$ 0.0023 & 20.33\\
\noalign{\smallskip}
\hline 
\end{tabular}
\caption{Summary of the magnitude, colours and distances for our stars in the direction of Pal 5. The two stars at the bottom of the table are HB stars in the core of Pal 5.\label{stetson_hb}}
\end{table}

The QUEST survey of RR Lyrae stars (Vivas et al. 2004) found five
previously observed RR Lyrae stars in Pal 5 in the $V$ band; the data
for these stars are presented in Table \ref{rr_lyrae_quest}. The stars
in this table are 0.24 magnitudes fainter than our BHB stars,
suggesting they are more distant. Assuming an absolute magnitude of
the RR Lyraes of $M_V(RR)=1.112+0.214[\rm{Fe/H}]$ (as we discuss in
\S4.1), and an [Fe/H] = $-$1.4, then the mean distance of this sample
is 21.2 $\pm$ 0.93 kpc.

Two further Pal 5 HB stars are available online via the Canadian
Astronomy Data Centre(CADC)\footnote{The data is available at:
http://cadcwww.dao.nrc.ca/astrocat/ in the Stetson Standard Fields}
which are also observed in the SDSS. Like the QUEST RR Lyrae stars
these stars are more than 0.2 magnitudes fainter than the those
observed in this paper. Table \ref{stetson_hb} summarises the colours,
and $V$ magnitude of these two stars along with the four stars in our
study. If we again assume a metallicity of $-$1.4 then the stars s41
and s40 are at 20.0 and 20.3 kpc respectively. We note for
completeness that if the distances are derived using the SDSS colours
and magnitudes as in \S4.1 then we find the stars s41 and s40 are 20.9
and 21.9 kpc respectively.

Therefore, it seems unlikely that our BHB stars in the direction of
Pal 5 are part of the core of this cluster. However, the stars 70, 83
and 99 occupy a volume of space of 0.0004 cubic kpc, which is a space
density of around 8000 per cubic kpc.  Kinman et al. (1994; equation 12) find
a mean density of BHB stars at R$_{gal}$ = 14 kpc to be approximately
0.9 per cubic kpc. The three stars therefore appear to be in an
overdensity perhaps indicating that Pal 5 is extensive not only in the
plane of the sky but also in depth. Clearly more observations are
required to fully understand this structure and its possible relation
to Pal 5 and/or the Sgr stream.





\begin{figure*}
\centering{ \scalebox{0.5}{ \includegraphics*{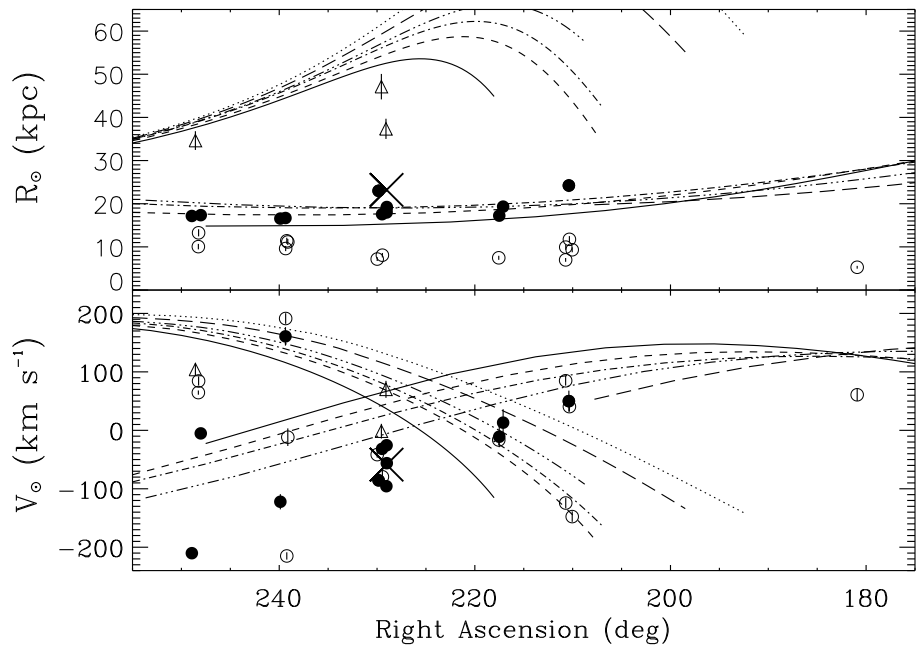}} }
\caption{{\em Upper:} Plot of R$_{\odot}$ versus RA for the BHB
stars. The BHB stars are separated in distance so that stars are at:
(i) 5 $<$ R$_{\odot}$ $<$ 14 kpc (open circles); (ii) 14 $<$ R$_{\odot}$ $<$ 25 kpc
(filled circles); and R$_{\odot}$ $>$ 25 (open triangles). The curves are taken
from  Mart{\'{\i}}nez-Delgado et al. (2004) and describe the axis ratio of the
density distribution, $q_d$. The values of $q_d$ range from 1.0 (dotted
line) to 0.1 (solid line) respectively. {\em Lower:} A plot of
V$_{\odot}$ versus RA for the three groups. The symbols and curves are
the same as those on the upper plot.}
\label{comp_delgado}
\end{figure*}

\section{Discussion}
Table \ref{data_sum} summarises the main observational results of the
paper; a sample of BHB stars with measured radial velocities in the
vicinity of the Sgr stream. We now investigate whether there is
evidence that these stars actually reside in this stream.

In Figure \ref{comp_delgado} we plot Heliocentric radial velocity and
distance versus RA for the 29 BHB stars. In Figure
\ref{comp_delgado}(upper) we split the sample into three groups of
R$_{\odot}$: stars at 5 $<$ R$_{\odot}$ $<$ 14 kpc (open circles); (ii) stars at
14 $<$ R$_{\odot}$ $<$ 25 kpc (filled circles); and stars at R$_{\odot}$ $>$ 25 (open
triangles). We overplot on this figure simulations of the tidal stream
of Sgr for various halo flatness taken from Mart{\'{\i}}nez-Delgado et
al. (2004). The curves on this plot are for values of the axis ratio
of the density distribution, $q_d$, which range from 0.1 to 1.0 (upper
to lower curves). However, the axis ratio of the potential, $q_p$, is
the quantity that determines the satellite orbit and hence the shape
of the potential. This parameter is a function of $q_d$ and
Galactocentric distance (see Fig. 11 in Mart{\'{\i}}nez-Delgado et
al. 2004) and ranges from 0.5 (oblate potentials) to 1.0 (spherical
potentials).

In Figure \ref{comp_delgado}(lower) we plot V$_{\odot}$ versus RA for
the same three groups of BHB stars. The plot reveals a possible
correlation between V$_{\odot}$ and RA which suggests that 10 of the
12 stars in the region 14 $<$ R$_{\odot}$ $<$ 25 kpc could be associated with a
stream. It is clear, however, that the velocity distribution of the
stars is systematically lower than any of the Mart{\'{\i}}nez-Delgado
et al. (2004) models presented here. There could be many reasons for
this. One plausible reason is that the Mart{\'{\i}}nez-Delgado model
considers only the last ($\sim 5\,$Gyr) orbit. As BHB stars are old,
any putative stream might feasibly belong to a later orbit that is not
considered in this model. In the next section we investigate the
evidence that this putative stream is plausibly associated with the
Sgr tidal debris.
\begin{figure*}
\centering{\scalebox{0.85}{ \includegraphics*[40,40][750,750]{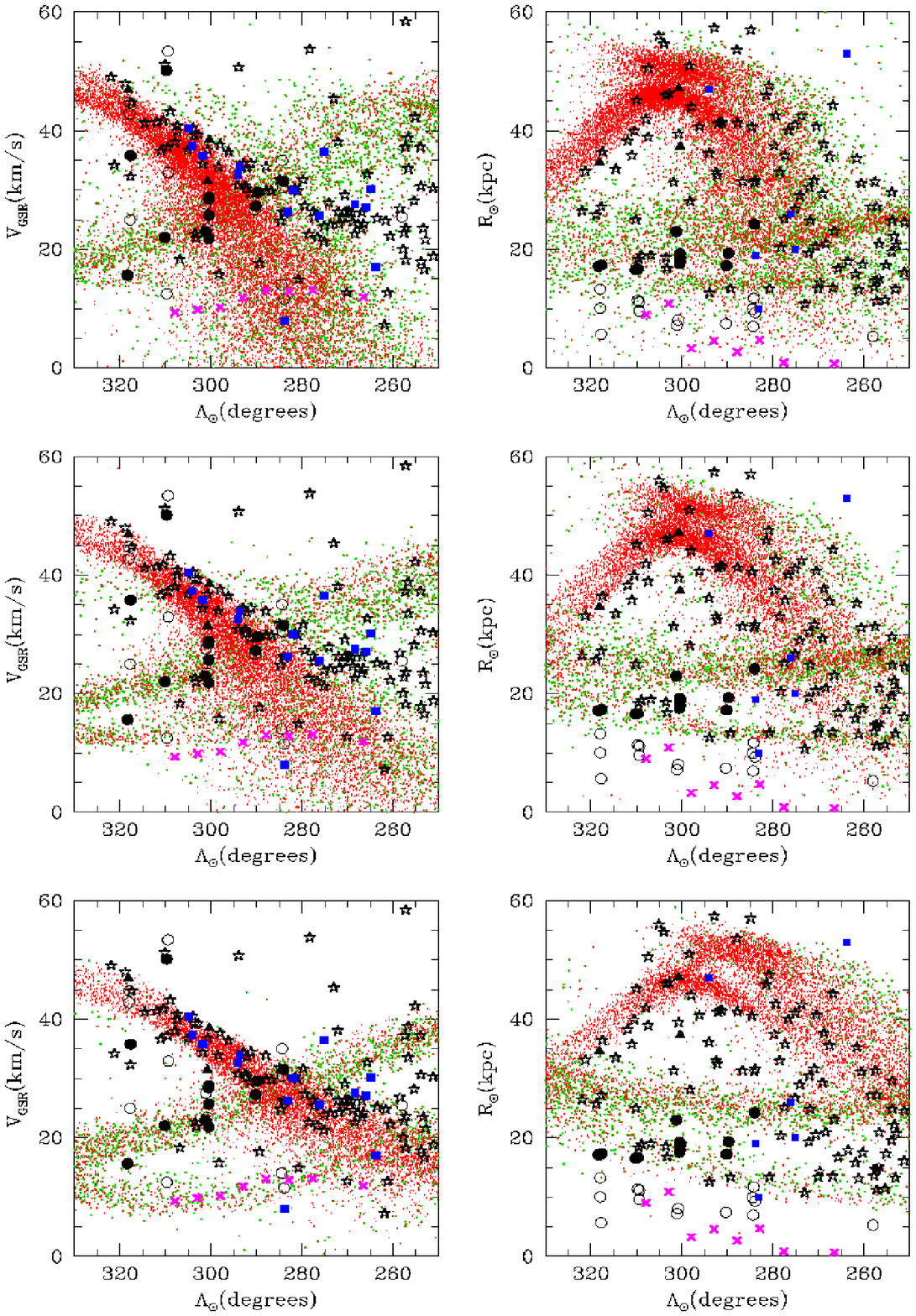}} }\caption{Distances and radial velocities of the best-fit simulations
of the Galactic halo potential created by Law, Johnston \& Majewski
(2005) for oblate potentials (top plots), spherical potentials (middle
plots) and prolate potentials (bottom plots). The bold dots (green)
are old debris stripped four orbits ago whilst faint dots (red) are
from earlier orbits. Overplotted are: BHB observations (as in
Fig. \ref{comp_delgado}, i.e. filled and open circles and filled
triangles); filled squares are carbon stars selected from Totten \&
Irwin (1998); open stars are M giants from Majewski et al. (2004) and
crosses (magenta) are K-giants from Kundu et al. (2002).\label{law05plot}}
\end{figure*}

\subsection{Comparisons with previous work}
As we briefly discuss in \S1 there have been numerous detections
of the Sgr tidal tail around the sky using tracers that map different
epochs, i.e orbits in which the debris was stripped from the
satellite. For example, Majewski et al. (2003, 2004) isolated M giants
in the Two Micron All-Sky Survey (2MASS) in order to map the position
and velocity distribution of tidal debris from Sgr around the entire
Galaxy. In Law, Johnston \& Majewski (2005; hereafter LJM05) these
stars are compared with numerous simulations of the tidal debris of
the Sgr satellite in differing Galactic potentials. We refer the
reader to this paper for details of the parameters used in this model. The
simulations are created using the relatively young M giants as the
principal observable. We note that BHB stars and carbon stars are
expected to trace older debris than M giants. We compare these models
with our BHB stars and other stellar types that are available.

Before we are able to compare our data with such models we need to
convert our coordinates to the system defined in Majewski et al.
(2003). In this coordinate system the zero plane of the latitude
coordinate $B_{\odot}$ coincides with the best-fit great circle
defined by the Sgr debris, as seen from the Sun; the longitudinal
coordinate $\Lambda_{\odot}$ is zero in the direction of the Sgr core
and increases along the Sgr trailing stream. Our sample resides in the
region $260^\circ< \Lambda_{\odot} <320^\circ$ and $-17^\circ <
B_{\odot} < 16^\circ$. This corresponds to a section of the leading
portion of the orbit which is particularly sensitive to the halo
shape. Indeed, M stars appear to strongly favour models with prolate
rather than oblate halos (Helmi 2004). In contrast, by fitting planes
to the leading and trailing debris and measuring the orbital
precession of the Sgr debris Johnston, Law \& Majewski (2005) find
oblate halos are favoured.

In Figure \ref{law05plot} we compare our BHB data with the best-fit
numerical simulations. Overplotted are various other tracers which we
discuss below. The M giants are represented by open stars. Figure
\ref{law05plot} illustrates the differences for the halo potential
models. We plot distances and radial velocities of the best-fit
simulations of the Galactic halo potential, which was created by LJM05 for oblate
(q=0.9, upper plots), spherical (q=1.0, middle plots) and prolate
(q=1.25, lower plots) potentials. The bold points (green) are old
debris stripped four orbits ago whilst faint points (red) are from
earlier orbits.

The BHB stars are shown as filled circles, open circles and filled
triangles as in Figure \ref{comp_delgado}. The filled circles
represent our putative stream. The BHB stars that are shown as open
circles do not appear to be associated with the models in either
velocity or distance. However, the 10 of 12 filled circles are
plausibly associated with the older debris both in velocity and
distance in the LJM05 model. More data is required to confirm whether
the three distant BHB stars (shown as triangles) are also associated
with a more distant younger part of the Sagittarius stream. They do not
show a remarkable metallicity which might suggest they are in an
earlier orbit of Sgr. 

Carbon stars (filled squares) are from the catalogue provided by
Totten \& Irwin (1998), confining ourselves to the
same $\Lambda_{\odot}$ ranges as the BHB data and to similar distances
(improved distance estimates are added from Totten, Irwin \&
Whitelock 2000). Some of the carbon stars (blue squares) do appear to
follow the M stars in velocity, however their distances do not appear to 
be associated at all. This discrepancy might be due to the uncertainty in
distance measurements for individual carbon stars. It is also plausible
that the carbon stars trace older debris than the M giants as they can
have larger ages. This sample however does not show any correlation
with the BHB stars.

The crosses in Figure \ref{law05plot} are eight metal poor K-giants
discovered by Kundu et al. (2002) to have coherent, smoothly varying
distances and radial velocities. The Kundu et al. (2002) K-giant stars
represent relatively old debris stripped from Sgr three pericentric
passages ago. The velocities of this sample fit the models reasonably
well particularly for the prolate potential shown in Figure
\ref{law05plot}(lower). However, the distances remain uncertain in all three models. 

In summary, the M stars and carbon stars predominantly map out the
earlier tidal debris. The velocity data for both tracers appears to
qualitatively fit the prolate models rather well (Figure
\ref{law05plot}, lower). However the distance information is less
clear, particularly for the carbon stars. The velocity information of
the metal poor K-giants again appears to fit the prolate models well but
again the distance information does not support this case. The BHB
data are too scarce in this study to distinguish competing models of
halo flatness despite being comparitively accurate standard
candles. We emphasise that these observations are over a section of
the leading part of the stream. Conclusive evidence for halo flatness
will probably require multi-epoch observational data from leading and
trailing debris.

\section{Summary}
We close with a summary of the main points in this paper. We have
presented the first results of a pilot study to trace out the Sgr
stream using BHB stars. Spectroscopy of the A-type stars, obtained
with the 2dF, produced a sample of 44 stars with data of suitable
quality for classification into the classes BHB and A/BS. The final
sample (Table 4) comprises 29 stars classified as BHB. The
heliocentric distances range from 5 to 47 kpc, with heliocentric
radial velocities accurate to 10 km s$^{-1}$, on average, and distance
errors $<10\%$.

We find 10 of 12 BHB stars at $14-25\,$kpc are plausibly tracing a
stream.  By comparing this data with models from LJM05, we find that
these stars may be associated with the older debris of Sgr both in
velocity and distance. Further observations along the entire Sgr orbit
both along the trailing and leading stream and at greater distances
are clearly required to trace the full extent of this structure on the
sky. A BHB analogue of the 2MASS survey would complement such surveys
by probing the older tidal debris, which may be crucial in
constraining the shape of the Galaxy halo (e.g. Helmi 2004). For the
BHB stars which reside in the putative stream, we find three of our
BHB stars in the direction of the globular cluster Palomar (Pal) 5
appear to be in a foreground overdensity. More observations
around these stars are required to establish any link to Pal 5 and/or
the Sgr stream itself. Despite these controversies BHB stars have the
distinct advantage of being more reliable standard candles than M
stars and carbon stars and so the distance estimates are more
precise. We emphasise observations of BHB stars have unlimited
potential at providing accurate velocity and distance information on
distant halo streams. The next generation multi-object spectrographs
provide an excellent opportunity to accurately trace the full extent
of such structures.

\section*{Acknowledgements} 
We would like to thank the referee for a number of helpful
comments. We thank David Law and David Mart{\'{\i}}nez-Delgado for
providing us with their models.  The Sgr coordinate system conversions
made use of code at:
http://www.astro.virginia.edu/$^\sim$srm4n/Sgr/code.html. We thank
Caroline Van Breukelen for useful comments. The authors acknowledge
PPARC for financial support. This paper uses observations made on the
Anglo Australian Telescope at Siding Springs using the 2dF instrument.
We made use of the SDSS online database.  Funding for the creation and
distribution of the SDSS Archive has been provided by the Alfred
P. Sloan Foundation, the Participating Institutions, the National
Aeronautics and Space Administration, the National Science Foundation,
the U.S. Department of Energy, the Japanese Monbukagakusho, and the
Max Planck Society.

\end{document}